\documentclass[letterpaper, 10 pt, conference]{ieeeconf}  
\usepackage{caption}

\IEEEoverridecommandlockouts                              

\usepackage{graphicx} 
\usepackage{subcaption}
\usepackage{multirow}
\usepackage{xcolor} 
\title{\LARGE \bf
Detecting Interbrain Synchronization in EEG Hyperscanning with MUSE-S EEG headband*}

\author{Tarmo Lipping$^{1}$, Ahmad Sharif$^{1}$, Matin Beiramvand$^{1}$ and Jari Turunen$^{1}$
\thanks{*This work was supported by the EU regional funds grant Well-being, Vitality and Smart Services Through Experience Production and Technology (EHEÄ)}
\thanks{$^{1}$All authors are with the Faculty of Information Technology and Computing Sciences, Pori Campus,
        Tampere University, Pohjoisranta 11A, 28100, Pori, Finland
        {\tt\small tarmo.lipping@tuni.fi}}%
}
\begin{document}
\maketitle
\thispagestyle{empty}
\pagestyle{empty}
\begin{abstract}
In this study preliminary results on the classification of EEG hyperscanning data acquired using MUSE-S consumer-level EEG headband are presented. Five pairs of subjects were involved and the recording protocol contained three two-person tetris game sessions alternating with relaxation periods. The data were segmented and ten spectral and cross-coherence features were calculated. The features were arranged into feature matrices and Convolutional Neural Network model was trained to discriminate between the relaxation and gaming. Two different feature sets -- the full set and a set containing only inter-subject cross-coherence features were tested. The results indicate that using the full feature set, relaxation and gaming periods were perfectly discriminated. Using only inter-subject cross-coherence features 94 \% and 79 \% classification accuracy for training and testing data was obtained, respectively.       
\end{abstract}

\section{INTRODUCTION}
Hyperscanning, a method increasingly used in social neuroscience, can be defined as the simultaneous measurement of brain activity from multiple subjects \cite{zamm_practical_2024}. In some contexts, hyperscanning studies may also be  referred to as two-person or multiperson neuroscience \cite{saul_two-person_2022}.
Hyperscanning may have various aims, such as studying the efficiency of teamwork \cite{reveille_using_2024}, investigating the neurophysiological basis of interactions between subjects in collaboration or cooperation, improving engagement in collaborative learning \cite{tan_is_2023} or relieving social anxiety. While hyperscanning can be performed using various brain imaging modalities such as functional Near-Infrared Spectroscopy (fNIRS), functional Magnetic Resonance Imaging (fMRI) or Mangnetoencephalography (MEG), Electroencephalogram (EEG) is the most feasible technique when considering real-life applications. 

Xu et al. evaluated the performance of various signal coupling analysis methods in EEG hyperscanning studies on simulated EEG data \cite{xu_evaluation_2024}. The setup considered by the authors involved dual-channel frontal lobe recording using a headband. The methods considered included Pearson correlation, Spearman's rank correlation, Phase Locking Value, Partial Directed Coherence, Spectral Coherence and Cross-correlation. In other hyperscanning studies methods such as Wavelet Transform Coherence \cite{tan_is_2023}, Bispectral analysis and Granger Causality \cite{hakim_quantification_2023} have also been used. The research on EEG hyperscanning techniques and respective analysis methods is still gaining popularity and no certain methodology has proved superior. 

Recent years have witnessed a boom of appearance to the market of consumer-oriented EEG devices. Devices such as Neurosky, EMOTIV Insight, EMOTIV Epoc or MUSE, just to mention a few, are used in increasingly wide range of real-life situations. We have recently showed that classification accuracies of around 90 \% can be achieved in discriminating between various levels of mental workload induced by the n-back memory game using a simple MUSE-S EEG headband \cite{beiramvand_assessment_2024}\cite{lipping_assessment_2024}. In the present study our aim was to investigate the capability of detecting interbrain synchronization using the same device. The study presented here is a preliminary investigation to determine the feasibility and specifications for a more extensive experimental study with similar setup.

\section{MATERIAL AND METHODS}
\subsection{Recording setup}
Five hyperscanning recordings were performed using MUSE-S (InteraXon Inc.) EEG headsets. The MUSE-S headset comes in a form of easy-to-use headband with textile electrodes and is thus very convenient in real-life applications. A total five volunteers, forming five two-person pairs, participated in this preliminary study. The participants were healthy adults and were selected from the members of our research team (1 female, 4 male). MUSE-S has four EEG channels -- AF7, AF8, TP1 and TP2 -- from which the AF channels were selected for the analysis. The TP channels were discarded due to excessive artifacts. 

\begin{figure*}[ht]
\centering
\includegraphics[width=\linewidth]{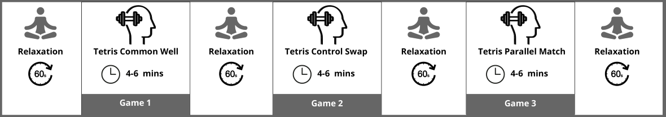}
\caption{Schematic of the recording protocol.}
\label{fig:protocol}
\end{figure*}

Two-person tetris game was used as the task in the hyperscanning recordings. The recordings involved three game sessions alternating with 1-minute relaxation periods (see Figure \ref{fig:protocol}). Three different versions of the game, available from the STEAM gaming platform were used. In the first gaming session -- the Common Well -- the participants cooperated in filling a common well with the tetris blocks. The second game session was Control Swap, in which the participants had separate wells and the control of the alternate tertris blocks was swapped so that if a participant dropped one block to the right well, his/her next block was allocated to the left well (and vice versa). In the third game session, both participants had their separate wells, posing a competitive setting. At the beginning of the recording and after each game session, the relaxation period was started by one participant pressing the 'x' key on the computer keyboard. An audio command instructed the participants to relax and keep their eyes closed until, after a 1-minute period, another audio command ended the relaxation period. 

\begin{figure}[ht]
\centering
\includegraphics[width=\linewidth]{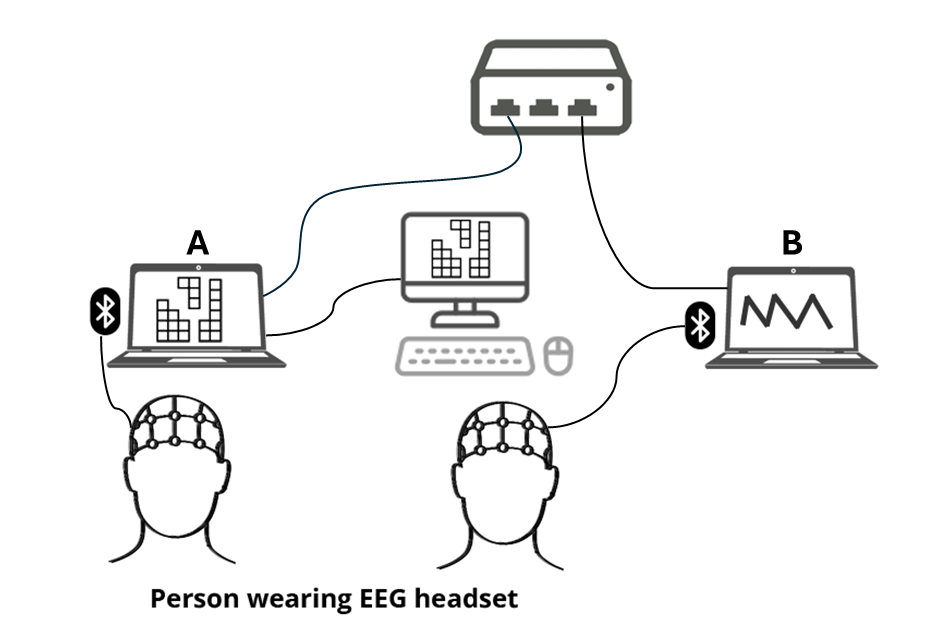}
\caption{Schematic of the recording setup. Laptop A runs the LabRecorder software, the tetris game, the script for recording keystrokes and the MuseLSL2 software to capture data from one participant. The laptop has a parallel monitor and keyboard for the other participant. Laptop B runs the MuseLSL2 software capturing the data from the other participant's MUSE-S device.}
\label{fig:setup}
\end{figure}

The data collection setup is presented in Figure \ref{fig:setup}. Two laptop computers were used to communicate with the MUSE-S EEG devices over Bluetooth. The MuseLSL2 software\footnote{https://github.com/DominiqueMakowski/MuseLSL2} was used on both laptops to produce the Lab Streaming Layer (LSL) data streams. One of the laptops was used to capture the streams over the LAN switch and store them in the .xdf format using the LabRecorder software. The particular laptop also ran the tetris game and a script to transform the keystrokes into an additional LSL stream. Keystrokes were then stored by LabRecorder together with the EEG data. The laptop running the game had duplicate monitors and keyboards for the two players. In local settings the LabRecorder software usually achieves a millisecond-level synchronization accuracy between the streams\footnote{https://github.com/labstreaminglayer/App-LabRecorder}. Approval for the study was granted by the Human Sciences Ethics Committee of Universities in Satakunta, Finland, no. 17.12.2024 

\subsection{Data Preprocessing and Feature Extraction}
The .xdf files were imported to MatLab using the \textsf{xdfread} tool for preprocessing and feature extraction. The data were first visualized to check for signal quality and to mark the start and end points of the relaxation periods and gaming sessions based on the recorded key strokes. The EEG signals, sampled at 256 Hz, were filtered using the 6th order Butterworth bandpass filter with cutoff frequencies 0.5 and 42 Hz. Subsequently, the signals of the two streams were resampled according to the timestamps in the .xdf files so that each stream had its samples exactly at the same time instances. Relaxation periods were marked by the specific keystroke (key 'x') and lasted for 1 minute. In one of the recordings, the last relaxation period after the third game was missing due to technical problems. The duration of the gaming sessions varied: the shortest and longest sessions lasted 3 minutes 51 seconds and 6 minutes 25 seconds, respectively.

\begin{figure}[ht]
\centering
\begin{subfigure}[l]{0.26\textwidth}
\centering
\includegraphics[width=\textwidth]{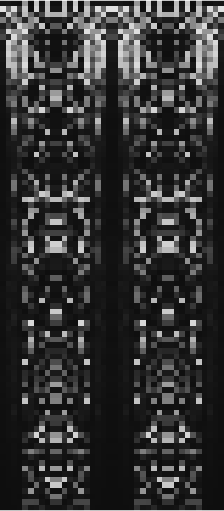} 
\end{subfigure}
\hspace{0.2cm}
\begin{subfigure}[r]{0.13\textwidth}
\centering
\includegraphics[width=\textwidth]{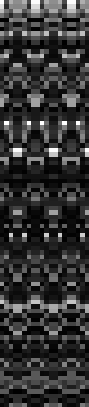} 
\end{subfigure}
\caption{Shape of the augmented feature matrix for a) the full feature set and b) the feature set containing only inter-stream cross-coherence features. Vertical axis denotes frequency from 0 to 45 Hz in 0.5 Hz steps; horizontal axis denotes the features as listed in the text.}
\label{fig:feature}
\end{figure}

For feature extraction, the relaxation periods and gaming sessions were divided into 10 second segments with 8 second overlap. Altogether, 475 relaxation segments (4 relaxation periods containing 25 segments each in all 5 recordings except that in one of the recordings the last relaxation period was missing) and 2161 gaming session segments were extracted. From each segment the following features were extracted:
\begin{enumerate}
    \item power spectral density of the four EEG signals (two channels and two streams) scaled to the range [0, 1] (4 features)
    \item magnitude squared coherence between the AF7 and AF8 channels of respective streams (2 features)
    \item magnitude squared coherence between the two channels of different streams (altogether 4 combinations, yielding 4 additional features).
\end{enumerate}
These basic features were selected as our aim was to test the feasibility of detecting interbrain synchrony in the particular setup rather than compare different feature sets. All the features were calculated within the frequency range from 0 to 45 Hz with 0.5 Hz step, forming a feature matrix of size $91 \times 10$. To better accommodate the next step, i.e., classification using the Convolutional Neural Network (CNN), the feature matrices were augmented by duplicating them 4 times with flipping. An augmented feature matrix is illustrated Figure \ref{fig:feature}a.

In addition to the full feature set, a subset containing only the inter-stream cross-coherence features (i.e., features listed in item 3 of the above list) was also used. The augmentation of the feature matrix was done in a similar manner as for the full feature set, except that the 4 features were duplicated (with flipping) 5 times forming a $91 \times 20$ feature matrix to be fed into the classification step (see Figure \ref{fig:feature}b).  

\subsection{Classification}

The augmented feature matrices were classified using the Convolutional Neural Network (CNN) classifier. CNNs have shown superior capabilities in finding patterns in data matrices and in case of a low number of layers, they can be trained using limited amount of training data. The network architecture was as follows:
\begin{itemize}
    \item $3 \times 3$ convolutional layer with 10 output channels and ReLu activation function
    \item $2 \times 2$ max-pooling layer
    \item $3 \times 3$ convolutional layer with 20 output channels and ReLu activation function
    \item dropout layer with $p = 0.5$
    \item fully connected layer with ReLu activation function
    \item fully connected layer with sigmoid activation function.
\end{itemize}
In the case of the full feature set, the batch size and learning rate were set to 32 and 0.00009, respectively while for the inter-stream cross-coherence features the respective values were 16 and 0.0002. The Adam optimization algorithm and binary cross-entropy loss function were used. Several alternative CNN architectures and hyperparameter settings were tested, however, it was found that the classification results were not very sensitive to the specific architecture and hyperparameter values and therefore no further optimization of these setting was performed in this preliminary study.  

70 \% of the data were randomly used for training and 30 \% for testing. Altogether 50 training runs, each containing 50 epochs, were performed. As the data are highly unbalanced, i.e., there are about 4.5 times more gaming segments than relaxation segments, for each run 475 feature matrices were randomly selected from the gaming data. 

\section{RESULTS AND DISCUSSION}
\begin{figure}[ht]
\centering
\includegraphics[width=\linewidth]{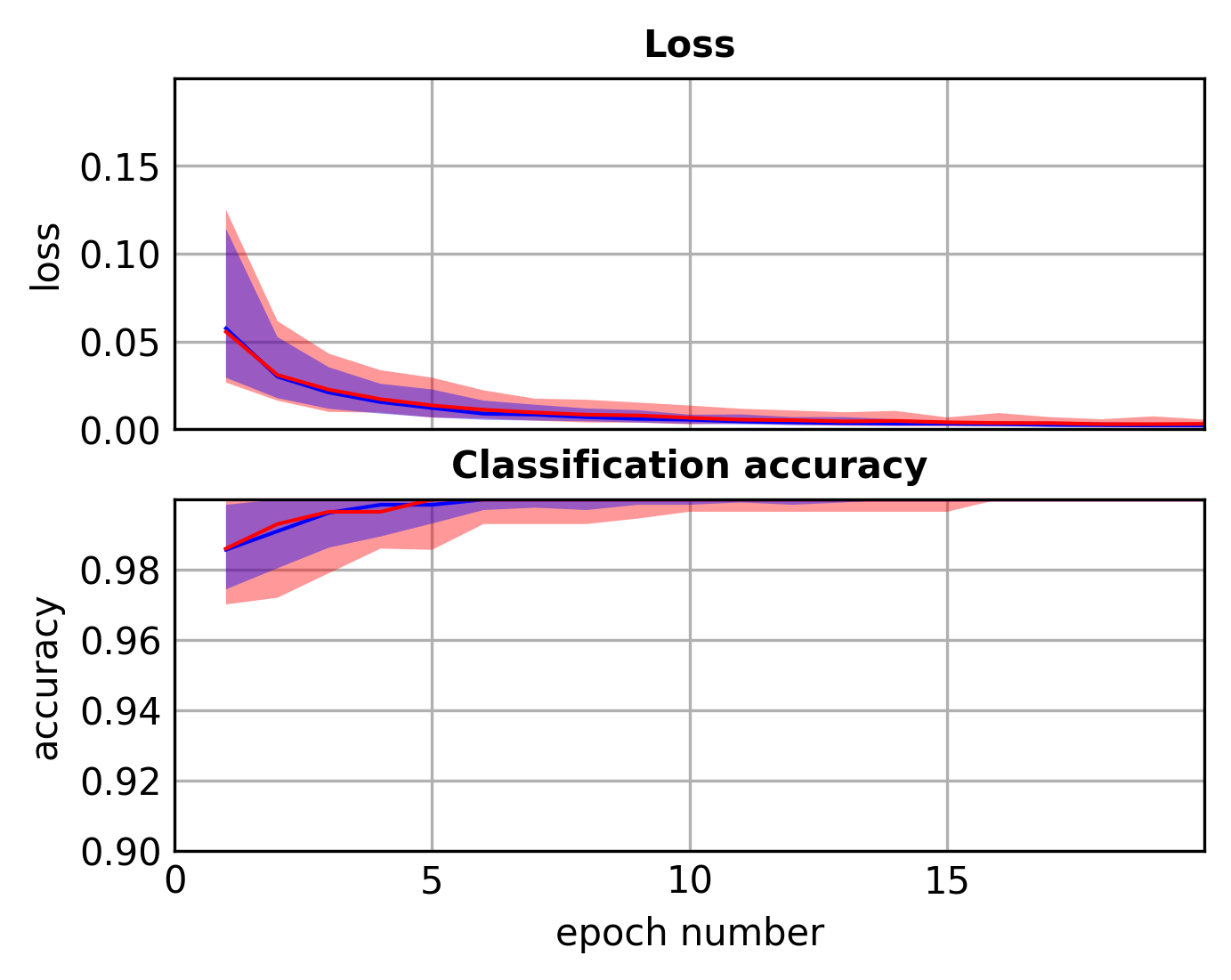}
\caption{Training loss and classification accuracy using the full feature set. The blue and red colors denote results obtained using training and testing data, respectively. The median and the range between 5 \% and 95 \% quantiles over the 50 runs are shown.}
\label{fig:random_full}
\end{figure}

\begin{figure}[ht]
\centering
\includegraphics[width=\linewidth]{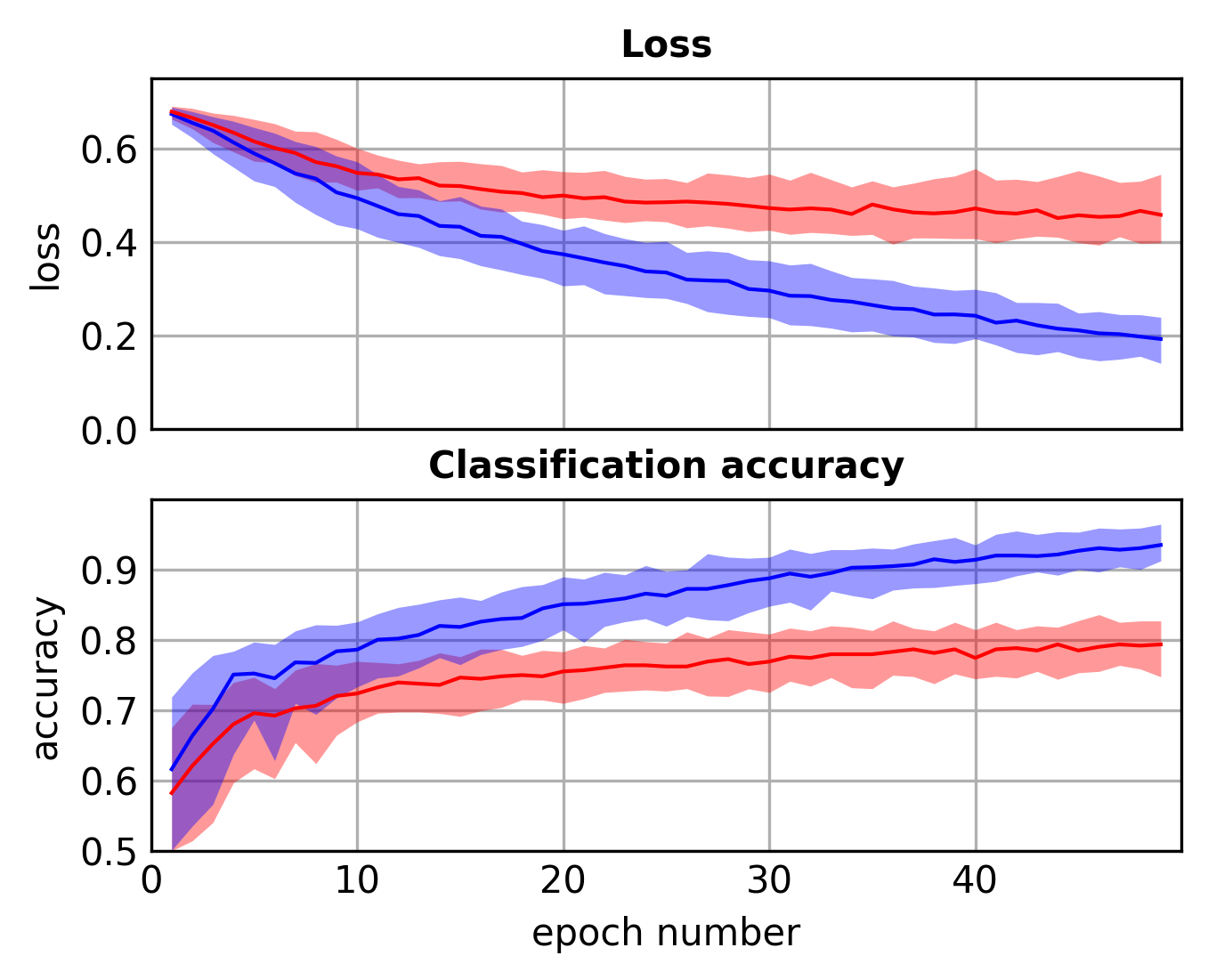}
\caption{Training loss and classification accuracy using only the inter-stream cross-coherence features. The blue and red colors denote results obtained using training and testing data, respectively. The median and the range between 5 \% and 95 \% quantiles over the 50 runs are shown.}
\label{fig:random_coh}
\end{figure}

Training loss and classification accuracy are presented in Figures \ref{fig:random_full} and \ref{fig:random_coh} for the two feature sets. Only 20 epochs are shown for the full feature set as in this case the CNN learns to discriminate between the relaxation and gaming basically during the first couple of epochs (note the different y-axis scales in the figures). This was achieved despite the very low learning rate (0.00009). In Table \ref{results} the classification accuracy after 50 epochs of training, aggregated over 50 runs is indicated for the two feature sets.

\begin{table*}[ht]
\normalsize
\caption{Classification accuracy after 50 epochs of training. Median $\pm$ Inter-Quartile Range (IQR) are indicated over the test runs.}
\label{results}
\begin{center}
\begin{tabular}{|l|c|c|}
\hline \noalign{\smallskip}
\multirow{2}{*}{\textbf{Feature set}} & \textbf{Training accuracy}  & \textbf{Testing accuracy} \\
& (median $\pm$ IQR) & (median $\pm$ IQR) \\
\noalign{\smallskip} \hline
\hline \noalign{\smallskip}
\textbf{Full feature set} & 1.00 $\pm$ 0.00 & 1.00 $\pm$ 0.00 \\
\noalign{\smallskip} \hline \noalign{\smallskip}
\textbf{Inter-stream coherence only} & 0.94 $\pm$ 0.02 & 0.79 $\pm$ 0.02 \\
\noalign{\smallskip} \hline 
\end{tabular}
\end{center}
\end{table*}

\addtolength{\textheight}{-1.5cm}

Our results indicate that when using power spectral density and intra-subject cross-coherence features, the CNN had no difficulty in discriminating between the relaxation and gaming. This could be expected as in addition to the mental activity, the two conditions also differed in that during relaxation the eyes of the subjects were closed. The nearly perfect classification is still noteworthy as the data were acquired using the consumer-oriented wireless EEG device. 

When only inter-stream (i.e., inter-brain) cross-coherence features were used, the learning process was significantly slower despite the higher learning rate (0.0002). The learning process follows the expected trend: further training improves the training accuracy (reaching 94 \% after 50 epochs) while the testing accuracy gets limited at about 75--80 \% after 20 epochs. It is clear that with only 5 subjects and random cross-validation, there is some leakage of information between the subjects. However, obtaining 79 \% accuracy using only inter-brain features and consumer-oriented EEG devices can be considered noteworthy.



\section{CONCLUSIONS}

In this study a preliminary analysis of five hyperscanning EEG recordings was performed. The CNN model was used to discriminate between relaxation and gaming using the EEG data described by the power spectral density and cross-coherence features. Three versions of the two-player tetris game were used. The main conclusions from the study can be summarized as:
\begin{itemize}
    \item when using spectral and cross-coherence features the relaxation and gaming states can easily be discriminated from one another using low-complexity consumer-oriented EEG headband with prefrontal channels
    \item when using only inter-subject cross-coherence features the two brain states could still be discriminated indicating that the EEG headband can detect some interaction between the subjects' brain activity in hyperscanning setting
    \item it is worth continuing the research by performing a more extensive series of hyperscanning recordings. 
\end{itemize}  

Based on these preliminary results, we are currently carrying out a more extensive recording series involving EEG hyperscanning using consumer devices. In the preliminary study we also tested if the different types of game could be discriminated by our model. The results were not promising in these regards. In the future experimental study we plan to better control the different collaborative and competitional tasks.



\bibliographystyle{ieeetr}
\bibliography{refs}

@inproceedings{lipping_assessment_2024,
	address = {Firenze, Italy},
	title = {Assessment of {Mental} {Workload} in {Real}-{Life} {Setup} using {EEG} {Synchronization} {Measures}},
	copyright = {https://doi.org/10.15223/policy-029},
	isbn = {9798350385823},
	url = {https://ieeexplore.ieee.org/document/10584156/},
	doi = {10.1109/MetroInd4.0IoT61288.2024.10584156},
	language = {en},
	urldate = {2025-02-05},
	booktitle = {2024 {IEEE} {International} {Workshop} on {Metrology} for {Industry} 4.0 \&amp; {IoT} ({MetroInd4}.0 \&amp; {IoT})},
	publisher = {IEEE},
	author = {Lipping, Tarmo and Beiramvand, Matin},
	month = may,
	year = {2024},
	pages = {412--416},
}

@article{xu_evaluation_2024,
	title = {An evaluation of inter-brain {EEG} coupling methods in hyperscanning studies},
	volume = {18},
	issn = {1871-4080, 1871-4099},
	url = {https://link.springer.com/10.1007/s11571-022-09911-1},
	doi = {10.1007/s11571-022-09911-1},
	language = {en},
	number = {1},
	urldate = {2025-02-05},
	journal = {Cognitive Neurodynamics},
	author = {Xu, Xiaomeng and Kong, Qiuyue and Zhang, Dan and Zhang, Yu},
	month = feb,
	year = {2024},
	pages = {67--83},
}

@article{reveille_using_2024,
	title = {Using interbrain synchrony to study teamwork: {A} systematic review and meta-analysis},
	volume = {159},
	issn = {01497634},
	shorttitle = {Using interbrain synchrony to study teamwork},
	url = {https://linkinghub.elsevier.com/retrieve/pii/S0149763424000629},
	doi = {10.1016/j.neubiorev.2024.105593},
	language = {en},
	urldate = {2025-02-05},
	journal = {Neuroscience \& Biobehavioral Reviews},
	author = {Réveillé, Coralie and Vergotte, Grégoire and Perrey, Stéphane and Bosselut, Grégoire},
	month = apr,
	year = {2024},
	pages = {105593},
}

@article{hakim_quantification_2023,
	title = {Quantification of inter-brain coupling: {A} review of current methods used in haemodynamic and electrophysiological hyperscanning studies},
	volume = {280},
	issn = {10538119},
	shorttitle = {Quantification of inter-brain coupling},
	url = {https://linkinghub.elsevier.com/retrieve/pii/S1053811923005050},
	doi = {10.1016/j.neuroimage.2023.120354},
	language = {en},
	urldate = {2025-02-05},
	journal = {NeuroImage},
	author = {Hakim, U and De Felice, S and Pinti, P and Zhang, X and Noah, J．A and Ono, Y and Burgess, P.W. and Hamilton, A and Hirsch, J and Tachtsidis, I},
	month = oct,
	year = {2023},
	pages = {120354},
}

@article{tan_is_2023,
	title = {Is neuroimaging ready for the classroom? {A} systematic review of hyperscanning studies in learning},
	volume = {281},
	issn = {10538119},
	shorttitle = {Is neuroimaging ready for the classroom?},
	url = {https://linkinghub.elsevier.com/retrieve/pii/S1053811923005189},
	doi = {10.1016/j.neuroimage.2023.120367},
	language = {en},
	urldate = {2025-02-05},
	journal = {NeuroImage},
	author = {Tan, S.H. Jessica and Wong, Jin Nen and Teo, Wei-Peng},
	month = nov,
	year = {2023},
	pages = {120367},
}

@article{beiramvand_assessment_2024,
	title = {Assessment of {Mental} {Workload} {Using} a {Transformer} {Network} and {Two} {Prefrontal} {EEG} {Channels}: {An} {Unparameterized} {Approach}},
	volume = {73},
	copyright = {https://creativecommons.org/licenses/by/4.0/legalcode},
	issn = {0018-9456, 1557-9662},
	shorttitle = {Assessment of {Mental} {Workload} {Using} a {Transformer} {Network} and {Two} {Prefrontal} {EEG} {Channels}},
	url = {https://ieeexplore.ieee.org/document/10510890/},
	doi = {10.1109/TIM.2024.3395312},
	language = {en},
	urldate = {2025-02-05},
	journal = {IEEE Transactions on Instrumentation and Measurement},
	author = {Beiramvand, Matin and Shahbakhti, Mohammad and Karttunen, Nina and Koivula, Reijo and Turunen, Jari and Lipping, Tarmo},
	year = {2024},
	pages = {1--10},
}

@article{saul_two-person_2022,
	title = {A {Two}-{Person} {Neuroscience} {Approach} for {Social} {Anxiety}: {A} {Paradigm} {With} {Interbrain} {Synchrony} and {Neurofeedback}},
	volume = {12},
	issn = {1664-1078},
	shorttitle = {A {Two}-{Person} {Neuroscience} {Approach} for {Social} {Anxiety}},
	url = {https://www.ncbi.nlm.nih.gov/pmc/articles/PMC8796854/},
	doi = {10.3389/fpsyg.2021.568921},
	urldate = {2025-01-12},
	journal = {Frontiers in Psychology},
	author = {Saul, Marcia A. and He, Xun and Black, Stuart and Charles, Fred},
	month = jan,
	year = {2022},
	pmid = {35095625},
	pmcid = {PMC8796854},
	pages = {568921},
}

@article{zamm_practical_2024,
	title = {A practical guide to {EEG} hyperscanning in joint action research: from motivation to implementation},
	volume = {19},
	copyright = {https://creativecommons.org/licenses/by/4.0/},
	issn = {1749-5016, 1749-5024},
	shorttitle = {A practical guide to {EEG} hyperscanning in joint action research},
	url = {https://academic.oup.com/scan/article/doi/10.1093/scan/nsae026/7641946},
	doi = {10.1093/scan/nsae026},
	language = {en},
	number = {1},
	urldate = {2025-01-12},
	journal = {Social Cognitive and Affective Neuroscience},
	author = {Zamm, Anna and Loehr, Janeen D and Vesper, Cordula and Konvalinka, Ivana and Kappel, Simon L and Heggli, Ole A and Vuust, Peter and Keller, Peter E},
	month = may,
	year = {2024},
	pages = {nsae026},
}

\end{document}